# Metallic *d*-wave altermagnetism in WFeB: a platform for electrically switchable perpendicular spin-splitter response

Eranga H. Gamage[1,2,#], Zhen Zhang[3,#], Subhadip Pradhan[4], Ajay Kumar[2], David R. Ramgern[1,2], V. Ovidiu Garlea[5], Yaroslav Mudryk[2], Saeed Kamali[6,7], Douglas Warnberg[8], Kirill D. Belashchenko[4], Vladimir Antropov[2,3,*], and Kirill Kovnir[1,2,*]

[1]*Department of Chemistry, Iowa State University, Ames, IA 50011, USA*
[2]*Ames National Laboratory, U.S. Department of Energy, Ames, IA 50011, USA*
[3]*Department of Physics and Astronomy, Iowa State University, Ames, IA 50011, USA*
[4]*Department of Physics and Astronomy and Nebraska Center for Materials and Nanoscience, University of Nebraska-Lincoln, Lincoln, Nebraska 68588, USA*
[5]*Neutron Scattering Division, Oak Ridge National Laboratory, Oak Ridge, Tennessee 37831, USA*
[6]*Department of Physics and Astronomy, Middle Tennessee State University, Murfreesboro, TN 37132, USA*
[7]*Mechanical, Aerospace & Biomedical Engineering Department, University of Tennessee Space Institute, Tullahoma, TN 37388, USA*
[8]*Research Engineering Group, University of Tennessee Space Institute, Tullahoma, TN 37388, United States*

[#]These authors contributed equally: Eranga H. Gamage, Zhen Zhang.
[*]Corresponding authors: Kirill Kovnir kovnir@iastate.edu, Vladimir Antropov antropov@iastate.edu

## Abstract

We report the synthesis and magnetic characterization of WFeB and identify it as a metallic *d*-wave altermagnet representative of a broader TiNiSi-type family. Neutron diffraction, Mössbauer spectroscopy, and magnetometry establish a collinear altermagnetic ordering confirmed by first-principles calculations. The electronic structure shows a nonrelativistic spin splitting of approximately 100 meV, but it also supports a strong spin-splitter transport response. This demonstrates that efficient spin-current generation can occur even with such modest band splitting. Symmetry analysis shows that selected film orientations permit deterministic switching of the Néel vector by current-induced staggered torques, enabling electrical control of a perpendicular spin-splitter response. These results establish WFeB and related TiNiSi-type antiferromagnets as a platform for electrically switchable charge-to-spin conversion driven by altermagnetic symmetry.

Altermagnetism, recently discovered as the third fundamental type of collinear magnetism after ferromagnetism and antiferromagnetism, simultaneously exhibits spin-split electronic bands characteristic of ferromagnets and zero net magnetization typical of antiferromagnets[1–7]. In conventional non-ferromagnetic materials, spin splitting of electronic bands typically relies on spin-orbit coupling (SOC). In contrast, altermagnets can exhibit spin splitting without SOC, purely as a consequence of symmetry constraints in the nonrelativistic limit. The potential spintronic applications of altermagnets arise from their ability to generate and detect spin currents, which can flow either parallel or perpendicular to the charge current. Among these, the pure transverse spin current, known as the spin-splitter effect[7–10], is particularly promising for achieving perpendicular magnetization switching in spin-orbit-torque-like device architectures, a key requirement for miniaturized magnetic memory devices[11].

In altermagnets, the spin-momentum distribution symmetry is often described using wave-type nomenclature, such as *d*-, *g*-, or *i*-wave[1]. The experimentally confirmed altermagnets, including MnTe[12] and CrSb[13], belong to the *g*-wave class. However, in the absence of SOC and strain, the nonrelativistic spin symmetry of the *g*-wave phase enforces a complete cancellation of the spin conductivity tensor, leading to unpolarized currents along any direction. Consequently, the practical spintronic utility of *g*-wave altermagnets is limited when SOC and strain are not involved. In contrast, *d*-wave altermagnets possess a spin-momentum symmetry that does not impose such restrictions, allowing for a robust spin-splitter effect in their metallic phases. While several insulating *d*-wave altermagnets such as rutile $MnF_2$ and $FeF_2$[14,15] have been experimentally confirmed, the search for metallic *d*-wave altermagnets has become a key thrust in altermagnetism research. The first proposed metallic *d*-wave altermagnet candidate was $RuO_2$, yet it has recently been decisively determined to be nonmagnetic in the bulk[16–18]. There is strong experimental evidence of metallic *d*-wave altermagnetism in $Mn_5Si_3$ thin films[19–23]. The existence of *d*-wave altermagnetism was suggested in the tetragonal inverse Lieb lattice-type oxychalcogenide family[24,25], but neutron diffraction measurements revealed magnetic unit cell doubling along the *c* axis in $KV_2Se_2O$[26]. This cell doubling is also predicted by first-principles calculations[27] in almost all vanadium oxychalcogenides, with the only exception of $CsV_2Te_2O$. Overall, the set of experimentally known metallic *d*-wave altermagnets remains extremely scarce.

All altermagnets (*d*/*g*/*i*-wave), as well as many noncollinear magnets, may exhibit anomalous Hall effect (AHE) depending on the orientation of the ordering vectors and the resulting magnetic point group. Recently, large AHE with only a small net magnetization was reported in NbMnP[28], TaMnP[29], and NbMnAs[30] compounds belonging to the orthorhombic TiNiSi structural prototype (*Pnma* space group). Based on neutron diffraction, it was reported that NbMnP is strongly noncollinear, with a combination of order parameters belonging to the $B_{3g}$ and $B_{2u}$ irreducible representations[31], while the chemically similar TaMnP is likely $B_{3g}$ with a weak DMI-induced magnetization, although the presence of a $B_{2u}$ component could not be ruled out[29]. The magnetic structure of NbMnAs, which orders above room temperature, has not been determined[30]. Multiple other antiferromagnetic (AFM) metals are known to have the TiNiSi structure[32]. For example, several *R*MnSi and *R*MnGe compounds (where *R* is a rare-earth element) are classified as collinear antiferromagnets. However, their magnetic ordering involves antiferromagnetically ordered Mn chains along the structural *b* axis[33,34]. This contrasts with the likely ferromagnetic (FM) chains found in TaMnP[29] and noncollinear chains present in NbMnP[31].

A straightforward symmetry analysis shows that the collinear magnetic ordering reported for TaMnP[29] belongs to the nontrivial spin point group (i.e., antisymmetric point group) $^2m^1m^2m$, which classifies them as *d*-wave altermagnets[1]. A combination of this *d*-wave spin-splitting type, metallicity, and reasonably high Néel temperatures makes the TiNiSi-type structure an excellent candidate for exploring new altermagnets. These materials could be used in spin-torque devices that exploit the spin-splitter effect to generate unconventional spin currents.

In this paper, we report, along with the synthesis, the magnetic characterization of the TiNiSi-type WFeB compound, which we identify as a metallic *d*-wave altermagnet. We further develop a symmetry-guided description of its altermagnetic band structure, spin transport, and surface-enabled switching responses, which is generalizable to other TiNiSi-based *d*-wave altermagnets. The WFeB compound is characterized by X-ray and neutron diffraction, magnetometry, $^{57}$Fe Mössbauer spectroscopy, and first-principles calculations. Theory and experiments consistently identify the same magnetic ordering in WFeB, whose symmetry implies *d*-wave altermagnetism. This ordering is characterized by FM zigzag Fe-chains with AFM order between the nearest chains. The Néel vector lies along the *c* axis at higher temperatures, allowing weak ferromagnetism, but reorients to the *b* axis below 150 K, where it is strictly compensated. Although the altermagnetic band splitting in WFeB is rather modest, approximately 100 meV near the Fermi level, it exhibits a strong altermagnetic spin-splitter effect with a large charge-to-spin conversion ratio. Moreover, a [001]-oriented WFeB layer is symmetry-allowed to generate an electrically switchable perpendicularly polarized spin-splitter current if it retains the high-temperature *c*-axis orientation of the Néel vector. This characteristic makes it a promising candidate for altermagnetic spin-torque devices. These symmetry-driven transport and switching responses apply broadly to TiNiSi-type antiferromagnets.

## Results and discussion
**Crystal structure and theoretical magnetic ordering**

WFeB crystallizes in the orthorhombic *Pnma* space group of the TiNiSi structure type with W, Fe, and B atoms occupying 4*c* sites (Fig. 1a)[35]. Fe and B atoms form chains of edge-sharing $FeB_4$ tetrahedra that are arranged in an alternate fashion (Supplementary Fig. S1). Each chain shares corners with the surrounding chains. The Fe-B framework consists of $Fe_3B_3$ hexagons and $Fe_2B_2$ rhombi. Fe atoms create zigzag chain structures that extend along the [010] direction (Fig. 1b). The bond distance between Fe atoms along the chains measures 3.164(1) Å (3.160 Å in theory), while the zigzag steps are significantly shorter at 2.435(1) Å (2.427 Å in theory), and the nearest distance between neighboring zigzag chains is significantly longer at 3.932(1) Å (3.913 Å in theory). The unit cell parameters derived from the powder X-ray diffraction (PXRD) and neutron powder diffraction structure refinements are similar (Supplementary Table S1). No mixed occupancy on W or Fe sites was detected by the structural refinements.

To evaluate the theoretical magnetic ground state for WFeB, we considered the FM and various compensated collinear magnetic configurations shown in Figs. 1c-1f. The AFM-1 (Fig. 1e) and AFM-2 (Fig. 1f) states have zigzag Fe-chains with intrachain AFM interactions. The difference between them is whether each layer of Fe atoms parallel to the $ac$-plane has the same (AFM-1) or both (AFM-2) spin orientations. AFM-1 and AFM-2 have opposite-spin sublattices that can be transformed into each other by an inversion symmetry. Thus, they are categorized as conventional AFM orders[1,2]. Apart from them, another compensated collinear order (Fig. 1d) exhibits intrachain FM interactions and interchain AFM interactions. This order has opposite-spin sublattices that can be transformed into each other only by rotation or mirror symmetries. Therefore, it is an altermagnetic order[1,2]. First-principles self-consistent total energy calculations of these magnetic orders provide a reasonable evaluation for the magnetic ground state. The altermagnetic order (Fig. 1d) is obtained as the ground state (-31.8 meV/f.u. relative to the nonmagnetic state). The FM order has lower energy (-19.4 meV/f.u.) than those of the two conventional AFM orders (-2.7 and -0.4 meV/f.u.). Thus, this structure favors the FM zigzag Fe-chains over the AFM zigzag Fe-chains.

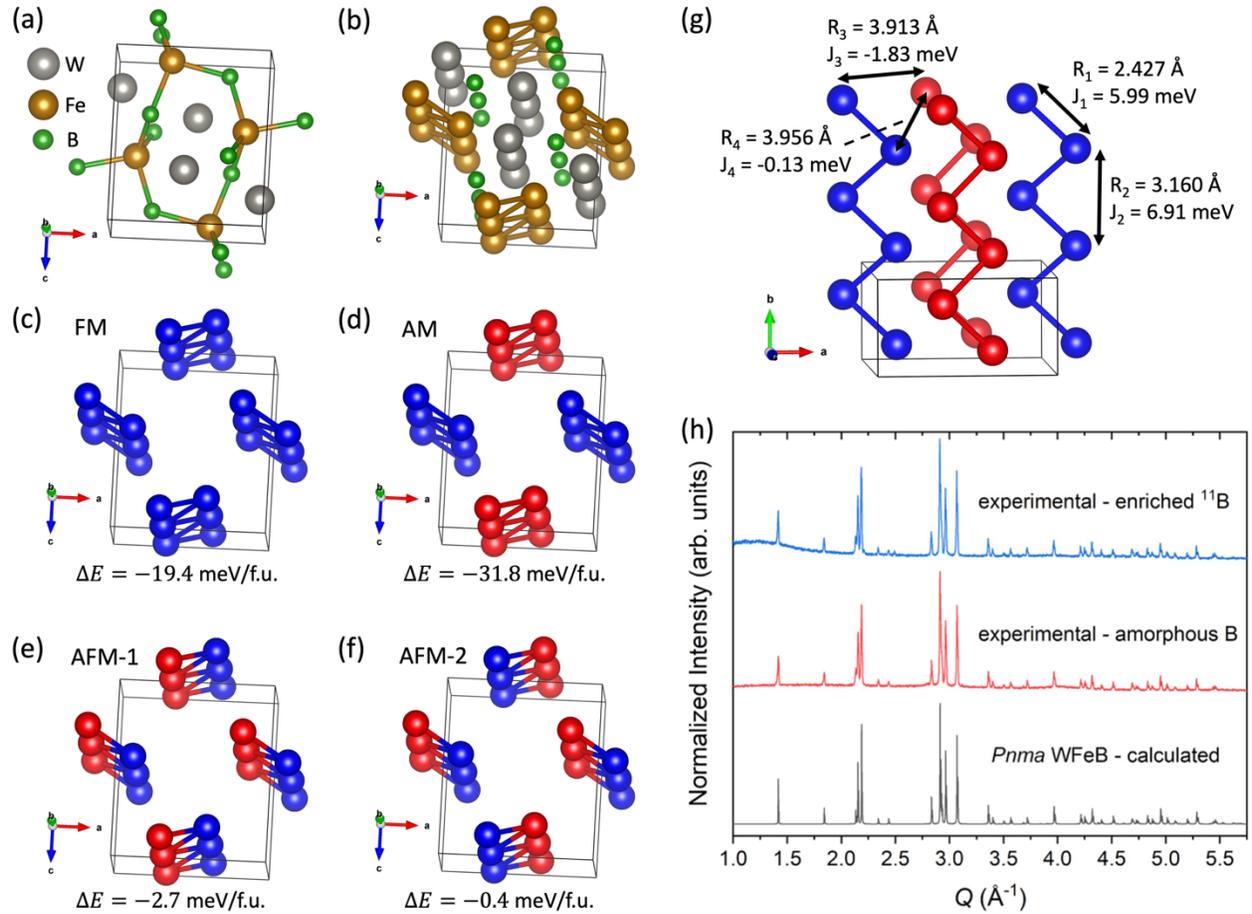

**Fig. 1. Crystal structure and theoretical magnetic ordering.** Crystal structure of orthorhombic WFeB with (a) Fe-B connectivity and interstitial W atoms, and (b) Fe-Fe zigzag chains propagating in the *b*-direction. (c)–(f) Different collinear magnetic configurations used for first-principles calculations. Blue and red spheres represent Fe atoms with opposite spins. W and B atoms are not shown. $\Delta E$ indicates the relative energy of the magnetic configuration with respect to the nonmagnetic configuration. (g) Magnetic exchange coupling parameters $J_1 - J_4$ (meV). $R_1 - R_4$ (Å): theoretical distance between the given Fe neighbor. (h) Experimental and theoretical PXRD patterns of WFeB ($\lambda$ = 1.5418 Å) prepared by the iodine-assisted synthesis method.

Figure 1g shows the magnetic exchange coupling parameters between the first four nearest Fe neighbors. The parameters are defined by $E = -\sum_{ij} J_{ij}\, \hat{m}_i \hat{m}_j$, where $\hat{m}_i$ are unit vectors, and each atomic pair is counted twice. Magnetic exchange coupling calculations confirm our total energy prediction for the magnetic ground state, i.e., intrachain FM interactions ($J_1$ = 5.99 meV and $J_2$ = 6.91 meV) and interchain AFM interactions ($J_3$ = -1.83 meV and $J_4$ = -0.13 meV). Interactions between further neighbors become less significant than the ones shown. The Néel temperature $T_N$ is estimated to be 230 K within the mean-field approximation (MFA), $T_N = (2/3)J_0/k_B$, where $J_0 = \sum_j J_{ij}\hat{m}_i\hat{m}_j$ is the total magnetic stability parameter between a given atom and the rest of the crystal.

**Synthesis**

The experimental investigation of the WFeB phase has proven quite challenging due to the demanding synthesis conditions required, including high temperatures, elevated pressures, and prolonged reaction times (a brief synthesis background is provided in the Supplementary Information). Additionally, the magnetic characterization of WFeB is complicated by the presence of competing ternary phases, such as $W_2FeB_2$ and magnetic binary iron boride admixtures. It is crucial to accurately identify these impurity phases to understand their influence on the WFeB property measurements. We successfully synthesized crystalline WFeB samples via an iodine-assisted solid-state synthesis method previously employed for $FeMo_2B_2$-type ternary borides[36]. The high-quality crystalline samples enabled the identification of $Fe_2B$ admixture phase and allowed for the accounting of its effects.

The elemental ratio required for the synthesis is non-stoichiometric. An excess of Fe and B relative to the target stoichiometry was required to react with iodine and form iodides, which were removed by washing. Such iodides might be important intermediates to enhance the reactivity of elements. For neutron diffraction, an isotopically enriched sample containing [11]B was synthesized. $Fe_2B$ admixture was present in all synthesized samples. But it was not observed in the laboratory PXRD patterns using Cu-$K_\alpha$ radiation ($\lambda$ = 1.5418 Å), primarily due to the preferred orientation and Fe fluorescence interference (Fig. 1h). Nevertheless, $Fe_2B$ appeared in the high-resolution PXRD pattern (Supplementary Fig. S2) due to the lack of fluorescence and high spatial resolution, and in the neutron powder diffraction patterns due to the higher neutron scattering ability of boron (Supplementary Fig. S3). Neither neutron nor synchrotron X-ray Rietveld structural refinements indicate Fe/W site mixing; each site is occupied solely by a single element type.

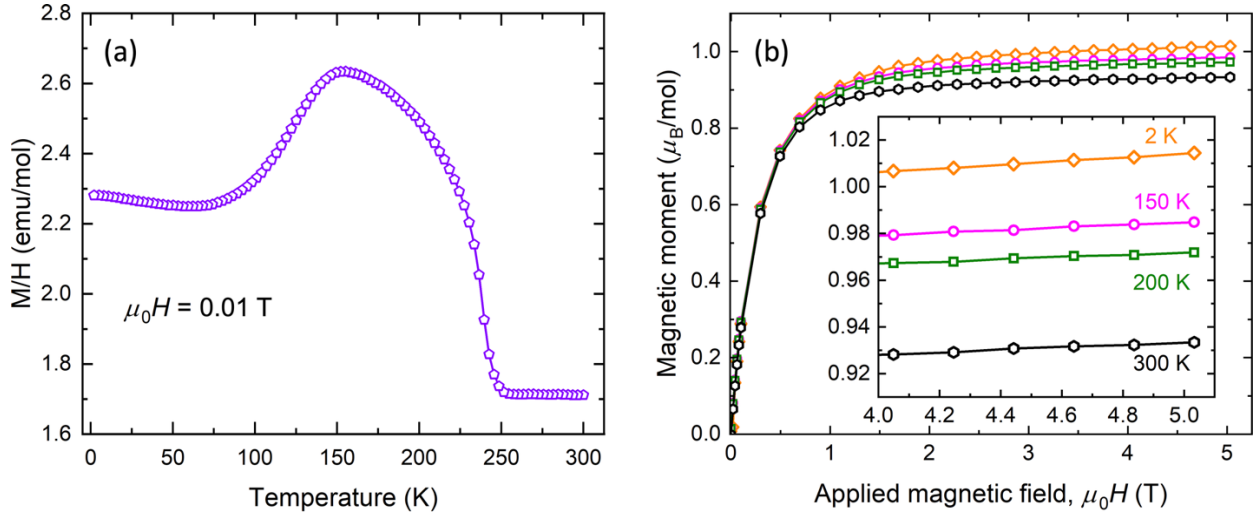

**Fig. 2. Magnetic measurements.** (a) *M/H* as a function of temperature measured under applied field of m $\mu_0 H$ = 0.01 T. (b) Isothermal field dependences of magnetization at different temperatures.

**SQUID magnetic measurements**

To investigate the magnetic behavior of WFeB, the magnetization as a function of temperature under applied magnetic fields of 100 and 1000 Oe was studied (Fig. 2a and Supplementary Fig. S4). The resulting magnetic susceptibility ($\chi$) curve does not follow the typical Curie-Weiss behavior. The presence of $Fe_2B$ admixture, a ferromagnet with high ordering temperature $T_C$ = 1013 K, resulted in an almost temperature-independent background signal. The zero-field cooled (ZFC) and field-cooled (FC) curves exhibit a significant bifurcation beginning at 240 K (Supplementary Fig. S4). This divergence of ZFC-FC curves indicates the onset of magnetic ordering. The FC curve shows a gradual upturn with a maximum $\chi$ value at 150 K. Following this peak, the magnetic moment drops to a minimum value at ~65 K. As the temperature decreases to 2 K, the moment slightly increases. The unusual behavior of magnetic susceptibility provides valuable insights into the presence of high magnetic anisotropy, which is associated with the preferred alignment of magnetic moments. The field dependence of magnetization was studied at different temperatures (Fig. 2b). A slight increase in the saturated magnetization with decreasing temperature may be due to the temperature dependence of the $Fe_2B$ moment or spin canting in WFeB. The FM component of WFeB, if present, is expected to cause a stronger increase in saturated magnetization. Since the magnetic signal is likely affected by trace magnetic impurities, we turned to $^{57}$Fe Mössbauer spectroscopy to verify that the observed behavior is due to the main WFeB phase.

**$^{57}$Fe Mössbauer spectroscopy**

$^{57}$Fe Mössbauer spectra were obtained at four temperatures, corresponding to the features observed in the magnetization curve (Supplementary Fig. S5). The summary of refined parameters is provided in Supplementary Table S2. The spectrum at 293 K can be fitted with three components: $Q_1$, $Q_2$, and $Q_3$ (Supplementary Fig. S5). The magnetically split $Q_1$ component with a chemical shift (δ) of 0.11 mm·s$^{-1}$ and a high magnetic hyperfine field ($B_{hf}$) of ~25 T is attributed to the Fe$_2$B admixture[37]. Fe$_2$B undergoes FM ordering at 1013 K[38], and hence it appears as a sextet at room temperature. The remaining major signal, $Q_2$, represents the Fe atoms in WFeB and appears as a non-magnetic doublet. This suggests that the Fe atoms in WFeB are not magnetically ordered at room temperature. This finding is consistent with the magnetic data, which also indicates an ordering event at 240 K. An additional doublet, $Q_3$, indicates a non-identified Fe-containing phase, which is probably amorphous since no other crystalline phases were detected by X-ray or neutron diffraction. This amorphous phase remains non-magnetic down to 6 K, yet its relative contribution increases with decreasing temperature, likely due to a larger recoil-free fraction than in the crystalline phases. This nonmagnetic phase should make a negligible contribution to the magnetic response of the sample.

Upon cooling, $Q_1$ sextet (Fe$_2$B) exhibits minor perturbations, such as the increase in chemical shift to ~0.1 mm·s$^{-1}$ due to the second-order Doppler effect. At 150 K, the $Q_2$ (WFeB) signal exhibits hyperfine splitting, which confirms magnetic ordering. $Q_2$ becomes an asymmetric sextet with a $B_{hf}$ of 6.4 T. As the temperature decreases further to 65 K and 6 K, the $B_{hf}$ of $Q_2$ slightly increases to 6.7–6.8 T. The $^{57}$Fe Mössbauer spectroscopy indicates that Fe atoms in WFeB exhibit magnetic ordering between 300 K and 150 K, with relatively minor changes upon further cooling down to base temperature. To further clarify this, we performed neutron diffraction studies.

**Neutron powder diffraction and magnetic structure solution**

Neutron-powder diffraction measurements were conducted using the POWGEN diffractometer at the Spallation Neutron Source (SNS), Oak Ridge National Laboratory. Data were acquired using ~1 g of WFeB samples synthesized using $^{11}$B. Measurements were performed at four different temperatures: 300 K, 150 K, 65 K, and 8 K (Fig. 3a). The pattern at room temperature did not show additional peaks apart from the nuclear peaks of WFeB and Fe$_2$B admixture (Supplementary Fig. S3). The 150 K data exhibit an additional magnetic peak that can be indexed as (100) using the same unit cell as the nuclear lattice. At 65 K, a new peak, which can be indexed as (001), starts to appear and becomes the dominant magnetic peak at 8 K base temperature (Figs. 3a and 3b). The intensity of (100) is relatively stable between 150 K and 65 K, remaining within the experimental limits. Magnetic ordering events reflected in the neutron diffraction patterns are consistent with the magnetic and Mössbauer measurements discussed above.

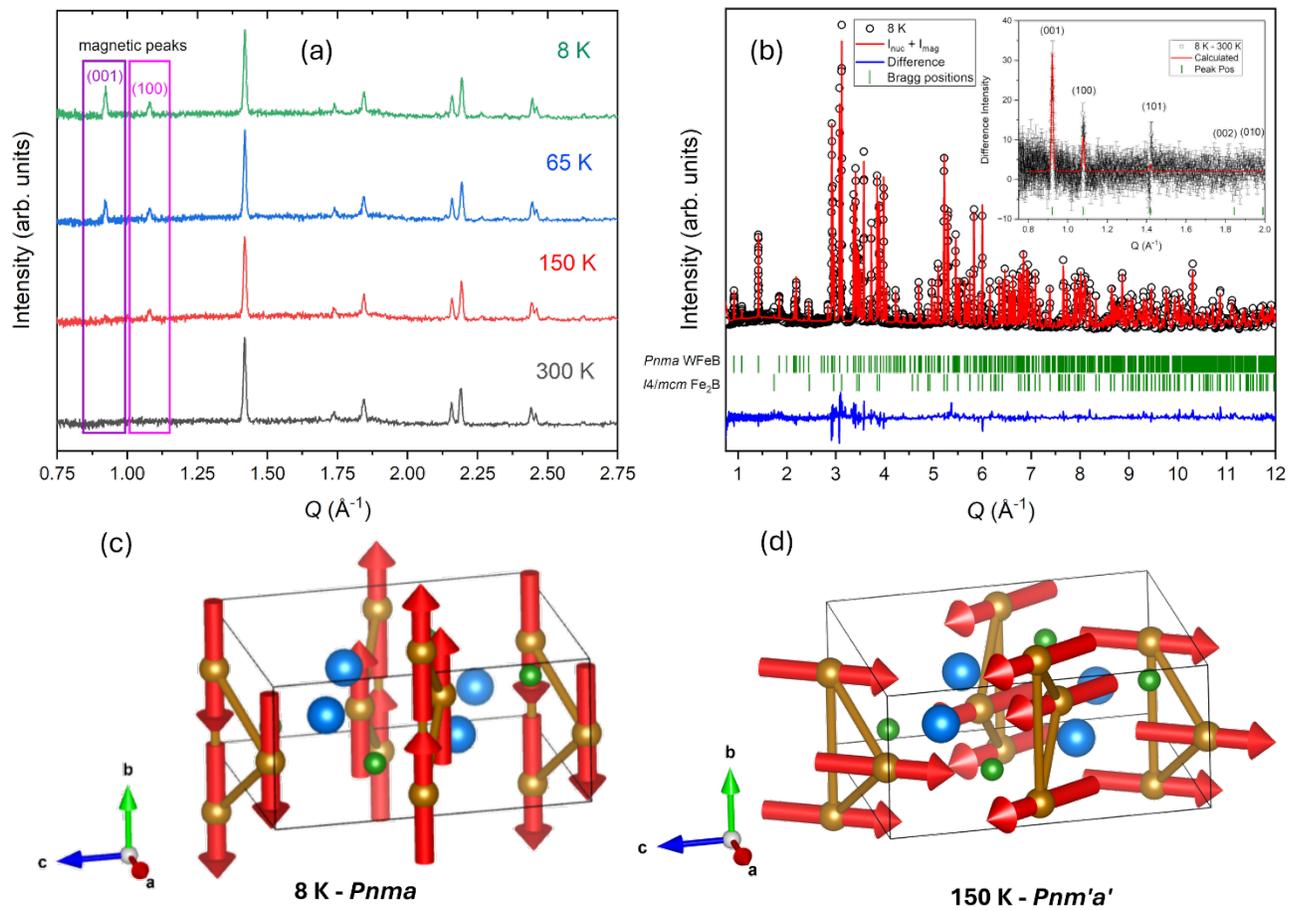

**Fig. 3. Neutron powder diffraction measurements.** (a) Neutron powder diffraction patterns of WFeB samples synthesized using $^{11}$B ($\lambda$ = 1.54 Å) measured at different temperatures. (b) Rietveld refinement performed on the 8 K data with nuclear and magnetic scattering included in the calculated profile. Inset: magnetic scattering obtained by subtracting 300 K data from 8 K data. Proposed magnetic structure model for WFeB at (c) 8 K, (d) 150 K.

The magnetic peaks observed in the 8 K pattern can be indexed using a propagation vector $\mathbf{k} = (0, 0, 0)$ for the average structure (Fig. 3b inset). Eight maximal magnetic space groups (MSG) that are compatible with the parent space group *Pnma* and the specified propagation vector $\mathbf{k} = (0, 0, 0)$ were identified using the symmetry analysis tools available at the Bilbao Crystallographic Server[39]. Out of these 8, only two, *Pnma* (#62.441) and *Pn'm'a'* (#62.449), allow both magnetic peaks (100) and (001). The best fit was obtained using the model described by *Pnma* (#62.441), with the magnetic moment aligned along the *b*-axis forming a collinear AFM arrangement (Fig. 3c and Supplementary Table S3). The collinear orientation is constrained by the magnetic symmetry that does not allow ordered components along the *a* and *c* directions. The refined moment amplitude is 0.8(1) $\mu_B$/Fe. It is important to note that the other MSG, *Pn'm'a'*, constrains the moment in the *ac*-plane ($M_x$, 0, $M_z$). However, this model produces a significant intensity at (010) that is not observed in the experimental data.

The fit of 150 K data, which shows only one magnetic peak at (100), is more complicated because multiple models could potentially fit the data. There are three candidate MSGs that would allow (100) but show no intensity at (001) (Supplementary Table S4). We successively tested: *Pn'm'a'* (#62.449) and *Pnm'a'* (#62.447), with moment along the *c*-axis, and *Pn'ma* (#62.443) with moment along the *b*-axis. Given the low intensities of the magnetic peaks, *R* factors are inadequate for selecting the best model. Instead, a graphical comparison should be used to evaluate how the models predict additional low-intensity peaks (see additional discussion and figures in the Supplementary Information). Magnetic structure of WFeB at 150 K is best described by the *Pnm'a'* (#62.447) model that allows for two moment components (Fig. 3d). It constrains a noncollinear moment to the *ac* plane, but with the *a*-axis projection being FM. The AFM component along the *c*-axis is approximately 0.7 $\mu_B$, based on the intensity of the (100) reflection. The *a*-axis component cannot be accurately determined because the FM component only slightly alternates the intensities of (201) and (102) nuclear reflections. It may vary from almost 0 to the upper limit of ~0.4 $\mu_B$, within a ±0.2 error. From Fig. 2b, the upper limit for the FM component is the difference between the saturated moments at 300 K and 200 K, 0.04 $\mu_B$. Taking the potential temperature-dependence of the FM $Fe_2B$ moment into account, the actual FM component of WFeB is smaller than 0.04 $\mu_B$. Supplementary Table S5 provides a description of the magnetic structure model at 150 K and its relationship with the parent paramagnetic structure.

The magnetic susceptibility measurements of the system are best described by the models developed by neutron diffraction. Magnetic ordering in Fe zig-zag chains is parallel, while intrachain arrangements are antiparallel, leading to an overall AFM nature. Upon cooling, the Fe spins order along [001] at 240 K. Further temperature decrease causes reorientation of spins along [010].

**$d$-wave altermagnetic order and band structure**

Both our calculations and measurements reveal the same magnetic ordering in WFeB: FM zigzag Fe chains with AFM order between the nearest chains. The theoretical magnetic moment on each Fe atom is 0.99 $\mu_B$ in the local density approximation (LDA), which is also in good agreement with the experimental values.

In the magnetic structure of WFeB (Fig. 4a), opposite-spin sublattices can be transformed into each other under nonrelativistic spin group symmetries, $[C_2||M_x\ t(0,½,½)]$ and $[C_2||M_z\ t(½,0,0)]$. Apart from a spin flip in spin space, the former indicates a $M_x$ mirror reflection followed by a translation along the diagonal of the *bc*-plane by half a lattice, while the latter indicates a $M_z$ mirror reflection followed by a translation along the *a*-axis by half a lattice, in real space. Correspondingly, in momentum space, the distribution of electronic spin also follows the $[C_2||M_x]$ and $[C_2||M_z]$ symmetries (Fig. 4b). The schematic spin-momentum locking pattern shown in Fig. 4b is characteristic of *d*-wave altermagnetism according to the classification of Ref.[1].

Such a *d*-wave spin-momentum pattern is further confirmed by the electronic band structure, which is shown in Fig. 4c. The altermagnetic band spin splitting, which is a nonrelativistic effect, is indicated by the opposite-spin blue and red bands. The magnitude of the altermagnetic band splitting is ~100 meV near the Fermi level. Bands with opposite spins exhibit spatial compensation, e.g., between Γ-K1 and Γ-K2, Γ-K3 and Γ-K4, Γ-K5 and Γ-K6, and between Γ-K7 and Γ-K8 under $[C_2||M_x]$, $[C_2||M_x]$, $[C_2||M_z]$, and $[C_2||M_z]$ symmetries, respectively. Such compensations for bands with opposite spins lead to zero net magnetization. Note that in conventional antiferromagnets, opposite-spin bands degenerate throughout the Brillouin zone. Therefore, the altermagnetic WFeB simultaneously exhibits zero net magnetization and band spin splitting. The partial electronic density of states on each atom is displayed in Supplementary Fig. S6.

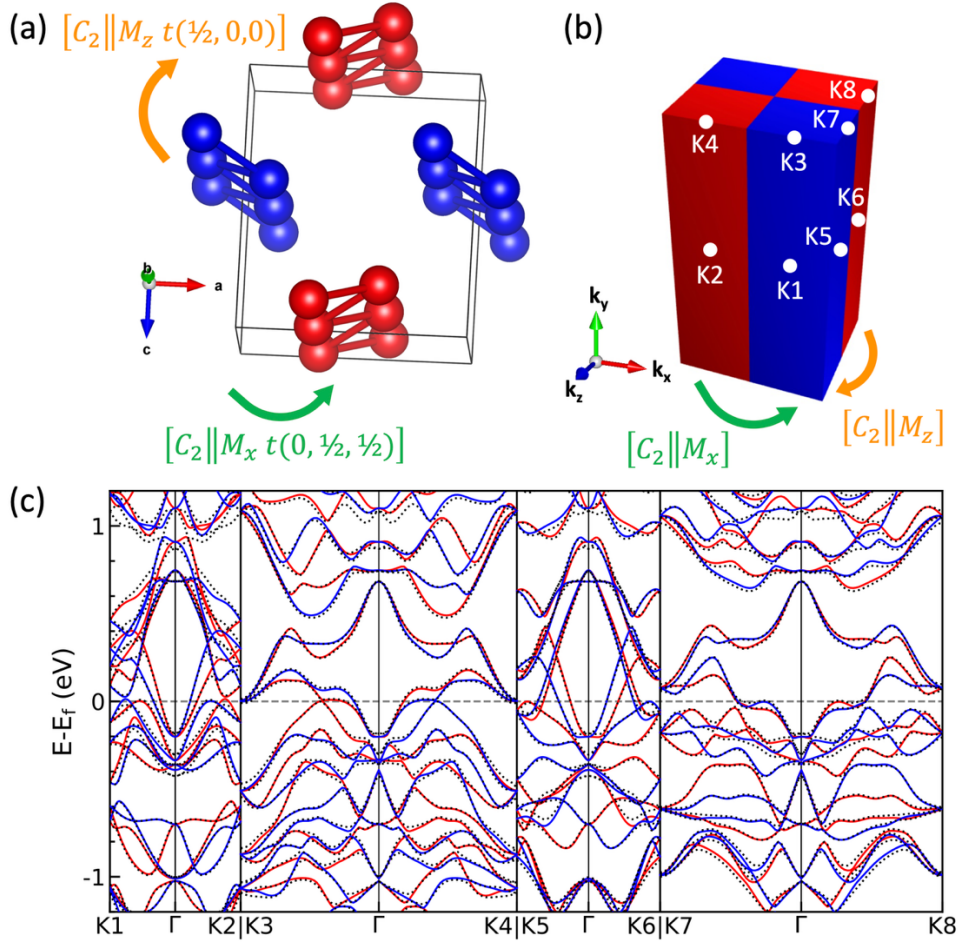

**Fig. 4. Altermagnetic order and band structure.** (a) Opposite-spin sublattices connected by nonrelativistic spin group symmetries. (b) Schematic spin-momentum locking corresponding to the magnetic structure. (c) Magnetic electronic structures along non-high-symmetry *k*-points: K1(1/4, 0, 1/2), K2(-1/4, 0, 1/2), K3(1/4, 1/2, 1/2), K4(-1/4, 1/2, 1/2), K5(1/2, 0, 1/4), K6(1/2, 0, -1/4), K7(1/2, 1/2, 1/4), K8(1/2, 1/2, -1/4). The blue and red curves show the spin-polarized electronic bands in the absence of SOC. Black dotted curves show the relativistic bands, including SOC, with the Néel vector along [010].

Since W is a heavy 5*d* element, relativistic effects, such as SOC, can give rise to additional band spin splitting. To quantify such effects, relativistic bands including the SOC are compared to the nonrelativistic spin-polarized bands in Fig. 4c. It is seen that the SOC effects are rather weak in WFeB, and the band spin splitting is primarily contributed by altermagnetism. Magnetic anisotropy calculations indicate that the ground state orientation of the Néel vector is along the [010] crystallographic direction, which is in good agreement with the experimental observation at 8 K (Fig. 3c). Magnetic states with the Néel vector along the [100] and [001] directions are higher in energy by 0.27 and 0.03 meV/Fe, respectively.

**Weak ferromagnetism**

As noted above, the *Pnm'a'* phase with the Néel order parameter oriented along the *c* axis is allowed by symmetry to have a magnetization along the *a* axis, which is realized via small spin canting induced by Dzyaloshinskii-Moriya interaction (DMI). We estimate the magnitude of this weak magnetization using VASP calculations for a series of configurations with the spins tilted from the *c* axis toward the *a* axis by a canting angle $\theta_c$. First, the total energy is calculated self-consistently with $\theta_c$ enforced using constraining fields and without SOC. The angular dependence $E_{ex}(\theta_c) = E_{tot}(\theta_c) - E_{tot}(0)$ represents the exchange contribution to the energy associated with the spin canting. Then, for each $\theta_c$, we calculate the DMI contribution $E_{\text{DMI}}(\theta_c)$ as $[E_b(\theta_c) - E_b(-\theta_c)]/2$, where $E_b$ is the band energy, by the generalized force theorem[40]. Here $E_b(\theta_c)$ and $E_b(-\theta_c)$ are calculated non-self-consistently, with SOC turned on and constraints turned off, using the same charge density, while the magnetization density for $-\theta_c$ is obtained by rotating the one for $\theta_c$ by 180° around the *c* axis. This method was previously used to calculate the transverse piezomagnetic effect in MnTe and CrSb[41]. The integration of $E_{\text{DMI}}$ was performed using a $14 \times 25 \times 12$ mesh in the full Brillouin zone, which, as we have checked, provided sufficient convergence.

The calculated $E_{ex}(\theta_c)$ and $E_{\text{DMI}}(\theta_c)$ are shown in Fig. 5. These two contributions are well fitted by even quartic and odd cubic polynomials, respectively. The equilibrium canting angle $\theta_{c0} = 0.23°$ is obtained by minimizing $E_{ex}(\theta_c) + E_{\text{DMI}}(\theta_c)$; the corresponding magnetization is only about 0.004 $\mu_B$/f.u.

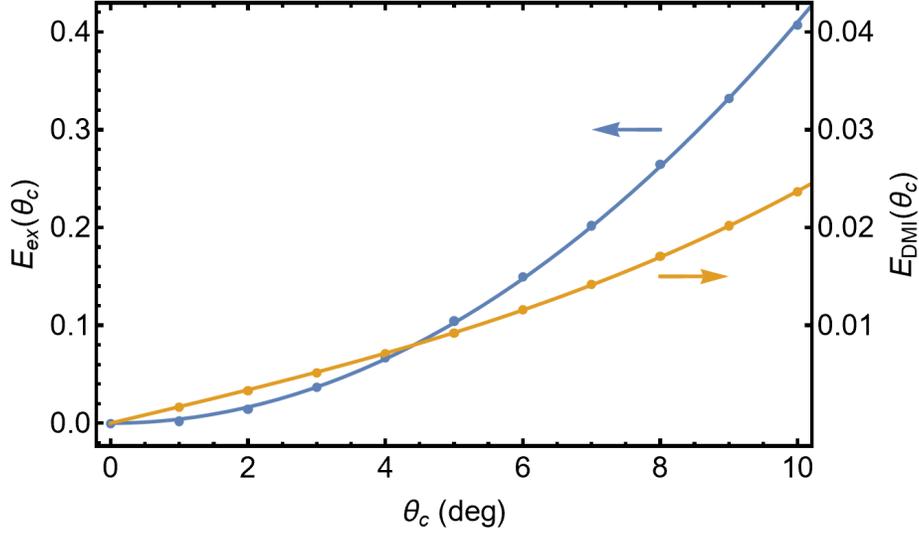

**Fig. 5. Weak ferromagnetism.** Exchange and DMI contributions to the canting energy in the *Pnm'a'* phase (units of μeV/f.u.).

**Anomalous Hall effect**

For WFeB, we calculated the AHE for all three principal orientations of the Néel vector **L**. Magnetic anisotropy calculations indicate that the ground state orientation of the Néel vector is along the [010] crystallographic direction. For this orientation, all components of the anomalous Hall conductivity tensor vanish by symmetry. However, if the Néel vector is oriented along [100] or [001], finite components $\sigma_{xy}^A$ or $\sigma_{yz}^A$ are allowed, respectively. The calculated dependence of these components on the chemical potential is shown in Fig. 6a. As is typical, the AHE exhibits strong energy dependence. At the Fermi level, both components have rather large magnitudes: $|\sigma_{xy}^A| = 328$ S/cm for **L** ∥ [100] and $|\sigma_{yz}^A| = 155$ S/cm for **L** ∥ [001].

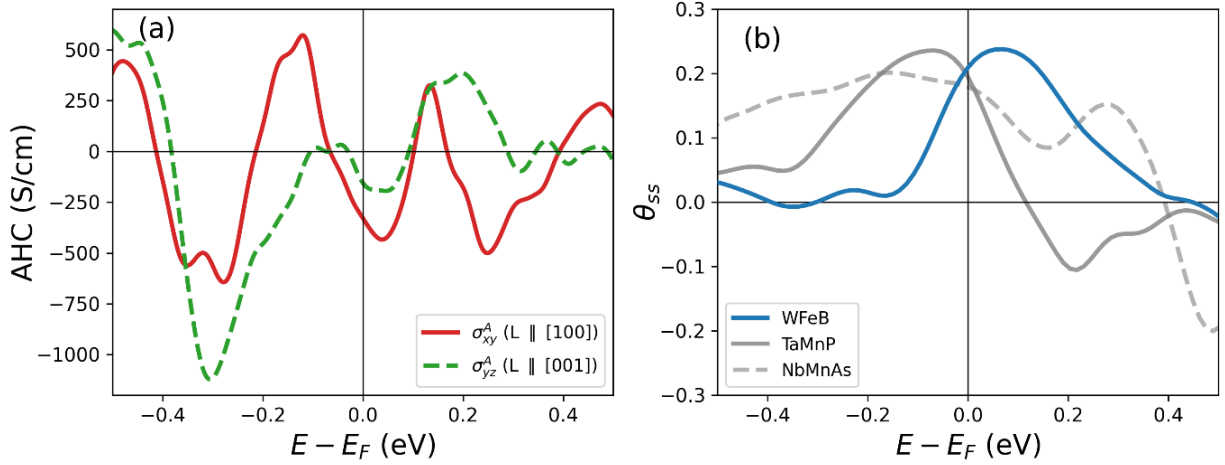

**Fig. 6. Transport properties.** (a) Anomalous Hall conductivity in WFeB: $\sigma_{xy}^A$ for **L** ∥ [100] and $\sigma_{yz}^A$ for **L** ∥ [001]. (b) Spin-splitter angle $\theta_{SS} = \sigma_{xz}^S/\sigma_{xx}$ in WFeB, TaMnP, and NbMnAs, assuming a collinear altermagnetic state.

**Spin-splitter effect**

While AHE is a spin-orbit-driven antisymmetric part of the charge conductivity tensor whose structure is determined by the magnetic point group, the spin-splitter effect is the symmetric part of the nonrelativistic spin conductivity controlled by the spin point group. In the chosen crystallographic setting, the nontrivial spin point group of TiNiSi-type altermagnets, including WFeB, is $^2m_x{}^1m_y{}^2m_z$, which is $d$-wave[1]. It allows one finite component $\sigma_{xz}^S$ of the symmetric nonrelativistic spin conductivity tensor $\sigma_{\alpha\beta}^S$, while the charge conductivity tensor $\sigma_{\alpha\beta}$ is diagonal. This means that a charge current that flows along the $x$ or $z$ axis is accompanied by a transverse spin current flowing perpendicular to the charge current in the $xz$ plane. The generation of such a transverse spin current in an altermagnet is called the spin-splitter effect[8], and it can potentially be used to switch the magnetization in an altermagnet/ferromagnet bilayer, similarly to the use of the spin-orbit torque due to the spin-Hall effect in a nonmagnet/ferromagnet bilayer[42]. In addition to the potentially larger charge-to-spin conversion ratio (which, in contrast to the spin Hall effect, does not rely on strong SOC in altermagnets), the spin polarization of the spin current in an altermagnet is parallel to the Néel vector and can have a component suitable for perpendicular magnetization switching, a highly desirable feature for magnetic random access memory applications[43].

To evaluate the nonrelativistic spin-splitter effect, we calculate the charge and spin conductivities within the Boltzmann approximation[44], assuming a constant relaxation time $\tau$ for all electronic states. The spin conductivity in this approximation reads

$$\sigma^s_{\alpha\beta} = \tau \sum_n \int v_{n\alpha} v_{n\beta} \sigma_n \frac{\partial f(E_n)}{\partial \mu} \frac{d^3 k}{(2\pi)^3}, \quad (1)$$

where $\mathbf{v}_n(\mathbf{k})$ is the group velocity, and $\sigma_n = \pm 1$ is the spin projection for the band $n$. The expression for the charge conductivity is the same, but without the $\sigma_n$ spin factor. The relaxation time cancels out in the spin-splitter ratio $\theta_{SS} = \sigma^s_{xz}/\sigma_{xx}$ or $\sigma^s_{xz}/\sigma_{zz}$, where the indices in the denominator correspond to the direction of the charge current. The tensors $\sigma^s_{\alpha\beta}/\tau$ and $\sigma_{\alpha\beta}/\tau$ were calculated using the smooth Fourier interpolation technique[45] implemented in the BoltzTrap2 package[46]. To facilitate the convergence of the Brillouin zone integration, we used the Fermi temperature of 300 K. Lowering the temperature to 100 K only adds some noise to the energy dependence of the spin conductivity (which is likely due to its imperfect convergence) without significantly affecting its value at the Fermi level.

Figure 6b shows the energy dependence of the spin-splitter angle $\theta_{SS} = \sigma^s_{xz}/\sigma_{xx}$ in WFeB, TaMnP, and NbMnAs, assuming the same collinear altermagnetic ordering. Because longitudinal conductivity in these materials is close to isotropic, the choice of the charge current direction in $\theta_{SS}$ is qualitatively immaterial. In all cases, the $\sigma^s_{xz}$ spin conductivity component maintains the same sign over a fairly broad window around the Fermi level, suggesting it should be robust against moderate doping and disorder. Near the Fermi level, the spin-splitter angle $\theta_{SS}$ is approximately 0.2 in all three compounds; this typical magnitude compares favorably with the spin Hall angle of order 0.1 for Pt, which is commonly used in spin-orbit torque devices[42].

While the AHE has been reported in related TiNiSi-type compounds[28–30], the spin-group-based classification of the collinear magnetic phase and the resulting altermagnetic spin-splitter response have not been analyzed. Our results show that WFeB and related collinear TiNiSi-type magnets possess *d*-wave altermagnetic symmetry and generically exhibit strong spin-splitter response despite a relatively modest altermagnetic band splitting.

**Switchability and electrically controlled spin current generation**

For the realization of perpendicular magnetization switching in an altermagnet/ferromagnet bilayer driven by the spin-splitter torque, three conditions need to be satisfied: (1) the Néel vector $\mathbf{L}$ should have a finite component along the interface normal direction **n**, (2) the flow direction of the generated spin-splitter current should have a finite component along **n**, and (3) the sample should have an unequal population of the two time-reversed magnetic domains. Let us consider whether these conditions may be met for some of the high symmetry [100], [010], or [001] film orientations of WFeB.

The third column of Table 1 lists the bulk spin current $j_\perp^{\text{SSE}}$ flowing perpendicular to the film plane due to the spin-splitter effect driven by an in-plane current. This current is determined by the symmetry of the bulk nonrelativistic spin conductivity tensor, which in turn is fixed by the $^2m_x{}^1m_y{}^2m_z$ nontrivial spin point group. We see that $j_\perp^{\text{SSE}}$ vanishes for the [010] film orientation but is finite for the [100] and [001] orientations. If **L** is perpendicular to the surface of one of these films, $j_\perp^{\text{SSE}}$ is not only finite but also has the favorable perpendicular spin orientation. In particular, if the experimentally determined bulk [001] orientation of **L** in the high-temperature phase of WFeB persists in a [001] film, this film should generate a perpendicular spin-splitter effect (Z-SSE) in the presence of an unequal population of **L** domains. The simplest way to achieve an unequal population of **L** domains in this phase is to apply a sufficiently strong external magnetic field along the [010] axis that will couple to its weak FM magnetization.

For device applications, field-free electric initialization of the single-domain magnetic state is preferable. As shown in Ref.[47], current-induced spin-orbit torques with the staggered polarization, which is required for deterministic electric switching of **L**, may be generated in some orientations of altermagnetic films even when such torques are forbidden in the bulk. The allowed macroscopic staggered torque components are determined by the surface antisymmetry point group of the film, which is a subgroup of the bulk antisymmetry point group that leaves the surface normal **n** invariant. The second column of Table 1 lists the corresponding electrically switchable components of **L** in the WFeB films, along with the required components of the driving electric field. In particular, the perpendicular $L_z$ component in the [001] film can be deterministically switched by the $E_y$ field component, resulting in an electrically switchable Z-SSE effect in such a film. Likewise, the $L_x$ component in a [100]-oriented film could be switched by $E_y$, again resulting in switchable Z-SSE, if the easy **L** axis happens to be along the [100] direction in such a film. The easy axis in a thin film may be different from its bulk orientation due to surface anisotropy or magnetoelastic contributions.

The fourth column of Table 1 indicates whether in-plane AHE is allowed for each film orientation, and which component of **L** is required to observe it. We see that AHE is not allowed at all for an [010] film, while for [100] and [001] films it requires a component of **L** that is not switchable by current alone.

**Table 1. Symmetry-controlled properties of thin films of a material belonging to the spin Laue group $^2m_x{}^1m_y{}^2m_z$ for high-symmetry growth orientations $\hat{n}$.** Second column: components of **L** that are deterministically switchable by current-induced staggered spin-orbit torque driven by the corresponding in-plane **E**-field components. $j_\perp^{SSE}$: non-relativistic spin-splitter current flowing parallel to $\hat{n}$. AHE: components of the anomalous Hall conductivity tensor allowed in the presence of the corresponding components of **L**. Last column: transport responses of the film that are deterministically switchable by an in-plane current in the presence of the corresponding component of **L**; it is assumed that **L** has only one nonzero component. Z-SSE: spin-splitter effect with both the flow and spin polarization directions perpendicular to the film, i.e., parallel to $\hat{n}$.

| $\hat{n}$ | Switchability of **L** | $j_\perp^{SSE}$ | AHE | Switchable responses |
|---|---|---|---|---|
| [010] | $L_x \leftarrow E_x, L_z \leftarrow E_z$ | — | — | — |
| [100] | $L_x \leftarrow E_y$ | $pE_z$ | $\sigma_{yz}^A \leftarrow L_z$ | $L_x \to$ Z-SSE |
| [001] | $L_z \leftarrow E_y$ | $pE_x$ | $\sigma_{xy}^A \leftarrow L_x$ | $L_z \to$ Z-SSE |

Summarizing this section, we find that [100]- and [001]-oriented films of WFeB with perpendicular easy axis satisfy the symmetry requirements for the generation of a Z-SSE. Moreover, their magnetic domain state, which controls the magnitude and sign of Z-SSE, can be deterministically initialized and switched by current-induced torques with the staggered polarization. The bulk easy axis in the high-temperature phase of WFeB already has the "matching" [001] orientation, as long as it is retained in the [001] film. These symmetry properties enable electrical control of the generated spin current and provide a route to efficient field-free switching of perpendicular magnetization in a spin-torque device.

**Summary**

Our systematic experimental measurements and first-principles calculations for the newly characterized WFeB identify the TiNiSi-type family of metallic *d*-wave altermagnets as a promising materials platform where altermagnetic symmetry enables strong charge-to-spin conversion and surface-enabled switching. WFeB has a magnetic ordering temperature of about 240 K and a spin-reorientation transition from *c*- to *b*-axis below 150 K. Despite a fairly modest altermagnetic band splitting, first-principles calculations predict a strong spin-splitter effect with a spin-splitter angle exceeding 0.2, along with a large AHE. This result demonstrates that sizable spin-splitter and anomalous Hall responses, which are the true figures of merit for key spintronic applications, can arise even for modest altermagnetic band splitting. With the *c*-axis Néel vector orientation, a [001]-oriented WFeB layer can serve as an efficient, electrically switchable source of perpendicularly polarized spin-splitter current (Z-SSE) for spin-torque devices.

## Methods

**Chemical materials**

Boron powder (Amorphous & crystalline, -325 mesh, 98%, Thermo Scientific), atomically enriched $^{11}$B (99%, 3M), W powder (99.9%, 1–5 micron, Alfa Aesar), Fe powder (0.4 mmol, <10 μm, Alfa Aesar, 99.5%), and $I_2$ crystals (99.999%, Sigma Aldrich) were used as received from vendors without any pre-treatment.

**Experimental synthesis**

Bulk WFeB powders were synthesized using an iodine-mediated solid-state reaction. W, Fe, and boron powders were loaded into a fused silica ampoule. The ampoule was placed in an ice bath before adding iodine crystals to minimize evaporation during sealing. Ampoules were sealed under vacuum using a hydrogen-oxygen torch. The sealed ampoules were wrapped in silica wool and heated in a muffle furnace to 1100 °C over 12 hours, held at that temperature for 12 hours, and then cooled to room temperature over 5 hours. After opening in a fume hood to let iodine vapors escape, the contents were treated with ~50 mL of 50% HCl for 3–5 hours to remove iodine-containing byproducts and residual Fe metal. The samples were then filtered, dried overnight, and stored in a desiccator, although they are air- and moisture-stable.

This method for the effective synthesis of ternary borides has been developed and optimized by us, and a detailed account is provided in Ref.[36]. WFeB was first synthesized in a reaction aimed at W$_2$FeB$_2$, but a power failure led the furnace to turn off mid-profile. The resulting product was identified as WFeB, and we realized that shorter annealing times favor metastable WFeB over the more stable W$_2$FeB$_2$. This phase was then pursued with additional synthesis attempts, followed by optimization. Due to the differences in reactivity of amorphous boron and atomically enriched boron, the loading ratios and ramping rates are slightly different in the synthesis schemes (Table 2). The samples synthesized with $^{11}$B contained more Fe$_2$B admixture than samples made with amorphous B (Fig. 1h). The latter was used for magnetic measurements.

**Table 2. Optimized reaction conditions for the synthesis of WFeB.** The end products were determined by *ex-situ* PXRD and confirmed by EDX. The reactions were conducted on a 200 mg scale.

| Boron source | Loading ratios* | Ramp time/dwell temp/ dwell time | Admixtures by PXRD |
|---|---|---|---|
| Amorphous natural abundance B | 1.5W+3.5Fe+2B+2I$_2$ | 10h/1100°C/12h | Fe$_2$B |
| Crystalline $^{11}$B enriched | 1.5W+4Fe+2B+2I$_2$ | 9h/1100°C/12h | Fe$_2$B |

Iodine-assisted solid-state synthesis yielded substantially large, block-like chunks visible to the naked eye (Supplementary Fig. S7a). The back-scattered electron images of the bulk sample showed well-defined small crystallites (Supplementary Fig. S7b). The big chunks were very hard to break and required significant effort. Reasonably sized crystals were screened using a single-crystal diffractometer. The diffraction pattern showed overlapping rings with smeared spots, suggesting the polycrystalline nature of the chunks. After additional effort, the chunks were broken down into very small crystallites and screened once more. However, even these resulted in fused rings, making it impossible to determine the unit cell and hindering accurate structural analysis. Therefore, structural refinement was conducted using Rietveld refinement of high-resolution PXRD data (Supplementary Fig. S2). The details, parameters, and results of the refinement are provided in Supplementary Table S1. The normalized EDX composition, W$_{1.03(1)}$Fe$_{0.97(3)}$, matches the refined compositions, indicating no Fe/W mixing. Lightweight boron was not quantified by EDX, but the qualitative presence of boron in every acquisition site was confirmed. Details of experimental characterization techniques used, PXRD, SEM/EDS, magnetometry, Mössbauer spectroscopy, and neutron diffraction, are provided in the Supplementary Information.

**First–principles calculations**

First-principles calculations were carried out using the VASP[48] package employing the projector-augmented-wave (PAW) method[49]. Structural optimizations were conducted with the generalized-gradient approximation (GGA) functional. Band structures, magnetic, and transport properties were computed with the LDA functional in the GGA-optimized structure. A $\Gamma$-centered $9 \times 16 \times 7$ mesh was used for Brillouin zone integrations, and the plane-wave kinetic-energy cutoff was set to 600 eV. The convergence threshold for electronic self-consistency was $10^{-5}$ eV, and the one for ionic relaxation was 0.01 eV Å$^{-1}$. SOC was included non-perturbatively in a subsequent calculation with the converged nonrelativistic charge density. Magnetic exchange coupling parameters were computed by using the TB2J package[50] based on localized orbitals obtained by the OpenMX package[51].

**Wannier downfolding and tight-binding calculations**

The first-principles band structure served as the starting point for constructing maximally localized Wannier functions. These atomic-orbital-like functions are derived from the Fe-$d$, W-$d$, and B-$p$ orbitals, using the WANNIER90 code[52]. The resulting tight-binding Hamiltonian in the Wannier basis faithfully reproduces the band structure in the $\pm 5$ eV energy window around the Fermi level, as shown in Supplementary Fig. S8. The Berry curvature was evaluated via the Kubo formula[53]:

$$\Omega_n^{\alpha\beta}(\mathbf{k}) = -2 \operatorname{Im} \sum_{m \neq n} \frac{\langle u_{n\mathbf{k}} | \hat{v}_\alpha | u_{m\mathbf{k}} \rangle \langle u_{m\mathbf{k}} | \hat{v}_\beta | u_{n\mathbf{k}} \rangle}{(E_{m\mathbf{k}} - E_{n\mathbf{k}})^2}, \qquad (2)$$

where $\hat{v}_\alpha = \frac{1}{\hbar} \partial \hat{H} / \partial k_\alpha$ is the velocity operator and $E_{n\mathbf{k}}$ are the band eigenvalues. The intrinsic anomalous Hall conductivity was obtained using WannierTools[54] by integrating the Berry curvature of the occupied bands:

$$\sigma_{\alpha\beta}^{\mathrm{AHE}} = -\frac{e^2}{\hbar} \sum_n^{\mathrm{occ}} \int_{\mathrm{BZ}} \frac{d^3 \mathbf{k}}{(2\pi)^3} \Omega_n^{\alpha\beta}(\mathbf{k}), \qquad (3)$$

using a $200 \times 200 \times 200$ $k$-point mesh.

**Acknowledgements**

We are thankful to Dr. Kevin Stone at BL2-1 of SLAC for his support with the rapid access mail-in proposal and the high-resolution PXRD data collection. We also thank Dr. Qiang Zhang for his support with the neutron diffraction experiment at POWGEN at SNS. The beamtime was allocated to POWGEN on proposal number IPTS-34092. This work was supported by the U.S. Department of Energy (DOE) Established Program to Stimulate Competitive Research (EPSCoR) Grant No. DE-SC0024284. The Ames National Laboratory is operated for the U.S. DOE by Iowa State University under Contract No. DE-AC02-07CH11358. Computations were performed at the High Performance Computing facility at Iowa State University and the Holland Computing Center at the University of Nebraska. Use of the Stanford Synchrotron Radiation Light Source, SLAC National Accelerator Laboratory, is supported by the U.S. DOE, Office of Science, Office of Basic Energy Sciences under Contract No. DE-AC02-76SF00515. A portion of this research used resources at the Spallation Neutron Source, DOE Office of Science User Facilities operated by Oak Ridge National Laboratory. This manuscript was authored by UT-Battelle, LLC under Contract No. DE-AC05-00OR22725 with the U.S. Department of Energy.

# Supplementary Information

## Metallic *d*-wave altermagnetism in WFeB: a platform for electrically switchable perpendicular spin-splitter response

Eranga H. Gamage[1,2,#], Zhen Zhang[3,#], Subhadip Pradhan[4], Ajay Kumar[2], David R. Ramgern[1,2], V. Ovidiu Garlea[5], Yaroslav Mudryk[2], Saeed Kamali[6,7], Douglas Warnberg[8], Kirill D. Belashchenko[4], Vladimir Antropov[2,3,*], and Kirill Kovnir[1,2,*]

[1]*Department of Chemistry, Iowa State University, Ames, IA 50011, USA*
[2]*Ames National Laboratory, U.S. Department of Energy, Ames, IA 50011, USA*
[3]*Department of Physics and Astronomy, Iowa State University, Ames, IA 50011, USA*
[4]*Department of Physics and Astronomy and Nebraska Center for Materials and Nanoscience, University of Nebraska-Lincoln, Lincoln, Nebraska 68588, USA*
[5]*Neutron Scattering Division, Oak Ridge National Laboratory, Oak Ridge, Tennessee 37831, USA*
[6]*Department of Physics and Astronomy, Middle Tennessee State University, Murfreesboro, TN 37132, USA*
[7]*Mechanical, Aerospace & Biomedical Engineering Department, University of Tennessee Space Institute, Tullahoma, TN 37388, USA*
[8]*Research Engineering Group, University of Tennessee Space Institute, Tullahoma, TN 37388, United States*
[#]These authors contributed equally: Eranga H. Gamage, Zhen Zhang.
[*]Corresponding authors: Kirill Kovnir kovnir@iastate.edu, Vladimir Antropov antropov@iastate.edu

**EXPERIMENTAL METHODS**

**Synthetic background**

Transition metal borides are extensively investigated for their mechanical and catalytic properties, both theoretically and experimentally[1–3]. In the 1960s, researchers sought to increase metal content in boron-rich compounds to reduce boron-boron interactions, leading to significant structural diversity as 3D frameworks transformed into 2D and 1D motifs. This exploration led to the development of ternary borides and the investigation of metal-rich compounds in ternary phase diagrams. Rieger, Nowotny, and Benesovsky first identified the MoCoB and WCoB alloys in {Mo,W}-Co-B phase space in 1965[4]. In 1966, Haschke et al. investigated the W-Fe-B chemical space at 1000 °C, and reported the WFeB ternary phase, along with the $W_2FeB_2$, $W_2FeB_4$, and $W_{7.5}Fe_{7.5}B_{11}$ phases[5]. Nevertheless, they noted that WFeB was not isotypic with MoCoB. Here, the synthesis was conducted by reacting cold-pressed powders in a tungsten tube furnace under purified argon or by annealing in sealed quartz tubes at 950–1100 °C temperatures. However, Jeitschko in 1967 synthesized WCoB and WFeB single crystals by annealing induction-melted powder pellets in sealed silica ampoules for 3 weeks at 950 °C and confirmed that WFeB is isotypic with MoCoB[6]. Kuz'ma and co-workers conducted the phase analyses of W-{Fe,Co}-B alloy compositions in 1968 and confirmed WFeB and WCoB to be isostructural to MoCoB[7]. They grew single crystals and established the TiNiSi-type orthorhombic *Pnma* crystal structure. Their method of synthesis involved sintering pressed powders in evacuated silica ampoules at 900 °C for 20 days.

In the 1980s, a study was conducted to increase the strength and hardness of binary iron borides by infiltrating them with Nb, Mo, and W metals[8]. High-temperature infiltration methods at 1180–1280 °C, performed under vacuum, were required to produce W-Fe-B alloys. Two decades later, Jasper et al. explored the isothermal section of the ternary phase diagram of W-Fe-B at 1050 °C and reported the successful synthesis of WFeB and $W_2FeB_2$ by long-term annealing of arc-melted or reaction-sintered alloys[9]. Over the last 5 years, extensive research has been conducted to analyze W-Fe-B alloy powders, particularly to investigate their mechanical properties. Reactive synthesis of ball-milled powders at 900–1150 °C under vacuum showed the gradual transformation of $W_xFe_yB_z$ ternary phase into $W_2FeB_2$ phase[10]. In a recent study, mixtures of Fe, W, and FeB powders were pressed under 300-500 MPa pressures, and different W-Fe-B alloys compositions were fabricated by pressure-free sintering at ~1300 °C to investigate the resulting mechanical properties[11–13]. Microstructural analysis revealed that the WFeB phase was absent. However, the formation of competing phases WFeB and $W_2FeB_2$ was observed in samples synthesized by aluminothermic self-propagating high-temperature synthesis[14]. Phase and microstructural analyses demonstrated that trace amounts of γ-Fe can promote iron diffusion into $W_2FeB_2$, leading to a higher iron-containing WFeB phase over time.

Achieving single-phase samples in the W-Fe-B system requires high temperatures, pressures, or prolonged annealing due to the refractory nature and high melting points of W (3422 °C), Fe (1538 °C), and B (2076 °C). Gibbs free energy calculations in the temperature range of 725–1225 °C have established that $W_2FeB_2$ is the most stable phase in the system, while WFeB is metastable[10,11]. This means that longer annealing can result in $W_2FeB_2$ rather than WFeB. Therefore, the synthesis of single-phase samples of WFeB free of $W_2FeB_2$ could be challenging. We report the synthesis of WFeB using a previously reported single-step iodine-assisted solid-state reaction completed in just 12 hours. Access to this metastable phase can be gained using both amorphous and atomically enriched crystalline boron.

**Experimental characterization**
Powder X-ray diffraction (PXRD) was performed using a Rigaku Miniflex 600 with Cu-$K_\alpha$ radiation ($\lambda$ = 1.5418 Å) and a Ni-$K_\beta$ filter. High-resolution PXRD experiments at room temperature were conducted at the beamline BL2-1 of SLAC National Accelerator Laboratory, Stanford Synchrotron Radiation Light Source (SSRL). The energy of the X-rays was 17 keV ($\lambda$ = 0.72964 Å). The samples were diluted with silica powder to minimize X-ray absorption ($\mu R$ ~1–2) and packed into Kapton capillaries of ~0.7 mm outer diameter and <2 mm height. $LaB_6$ was used as the calibration standard. Rietveld refinement was performed on PXRD data using the GSAS-II software package (version 5804)[15]. The background was first refined using the Chebyshev polynomial method, followed by unit cell refinement, sample displacement, size, microstrain, and preferred orientation. Next, the atomic coordinates and atomic displacement parameters (ADPs) for W and Fe were refined. Boron coordinates were not refined, and the ADPs were fixed to be close to ADP values refined for transition metals.

Elemental analysis of WFeB was conducted by energy dispersive X-ray spectroscopy (EDX) coupled with scanning electron microscopy (SEM) technique on an FEI Quanta 250 field emission-SEM with an EDX detector (Oxford X-Max 80, ThermoFischer Scientific, Inc., USA). The experiments were performed, and data were analyzed using Aztec software. Powder samples were sprinkled on carbon tape. Accelerating voltages of 5 kV and 15 kV were used to quantify the W: Fe ratio and detect admixtures in the samples.

Magnetic properties were investigated with a SQUID magnetometry (MPMS XL-7, Quantum Design). Temperature-dependent *dc* magnetic susceptibility measurements were conducted in both zero field cooled (ZFC) and field cooled (FC) modes in the temperature range 2–300 K under applied magnetic fields of 0.1 T and 0.01 T. In the ZFC mode, the sample was cooled in zero magnetic field, and the data were recorded upon warming in the presence of an applied magnetic field. In the FC mode, the sample was cooled in a constant magnetic field, and the data were recorded during warming in the same field. Isothermal magnetization measurements were performed at 2, 150, 200, and 300 K in a *dc* applied magnetic field ranging from 0 T to 5 T. The powders were placed in glass sample holders and protected from the atmosphere by Teflon caps.

$^{57}$Fe Mössbauer spectroscopic measurements were performed at 6, 65, 150, and 293 K using an MS4 spectrometer operating in the constant-acceleration mode in transmission geometry, with a 25 mCi $^{57}$Co in Rh source held at room temperature. For the low-temperature measurements, a Janis closed-cycle refrigerator was used. All centroid shifts δ, are given with respect to metallic α–iron measured at room temperature. All spectra were least-squares fitted by a Lorentzian

analysis using Recoil software[16] to extract hyperfine parameters, which are δ, quadrupole splitting/quadrupole shift ($\Delta E_Q/\varepsilon$), magnetic hyperfine field ($B_{hf}$), full-width at half-maximum ($\Gamma$), and intensity ($I$).

      Neutron-powder diffraction (NPD) was conducted at the powder diffractometer POWGEN, located at the Spallation Neutron Source, Oak Ridge National Laboratory. A polycrystalline sample was packed in a 6 mm Vanadium can with Cu gaskets. The neutron bank with a center wavelength of 1.5 Å was used to collect neutron data at four temperatures: 300 K, 150 K, 65 K, and 8 K. Rietveld refinements of the neutron diffraction data were performed using the FullProf Suite program[17]. Magnetic structure symmetry analysis was performed with computational tools at the Bilbao Crystallographic Server[18].

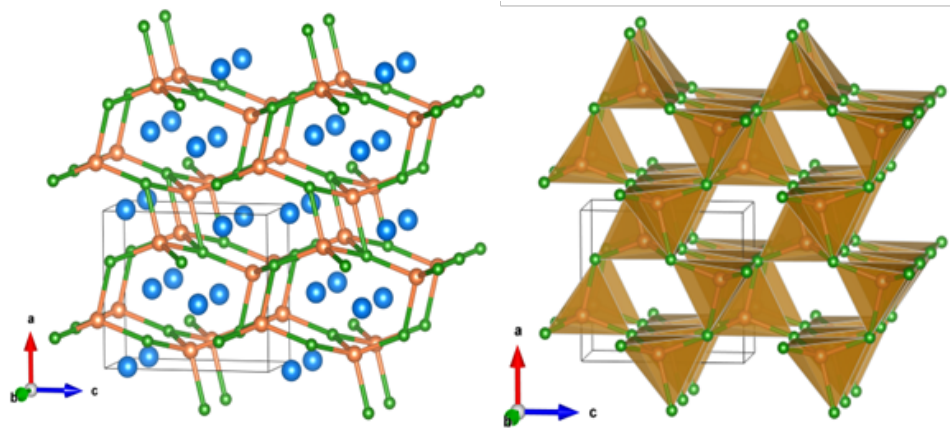

**Figure S1.** Crystal structure of orthorhombic WFeB emphasizing Fe-B connectivity. (a) Fe-B framework with interstitial W atoms. (b) Chains of edge and corner-sharing $FeB_4$ tetrahedra.

**Table S1.** Rietveld refinement parameters from HR-PXRD and NPD data for polycrystalline WFeB powders at 300 K.

| Phase | PXRD | NPD |
|---|---|---|
| wavelength (Å) | 0.72964 | time-of-flight |
| space group | *Pnma* (No. 62) | |
| $a$ (Å) | 5.824(1) | 5.826(1) |
| $b$ (Å) | 3.164(6) | 3.165(5) |
| $c$ (Å) | 6.816(1) | 6.817(2) |
| $V$ (Å$^3$) | 125.59(3) | 125.68(4) |
| $Z$ | 4 | 4 |
| $\rho$ (g•cm$^{-3}$) | 13.24 | 13.25 |
| no. of Bragg reflections | 408 | 2772 |
| no. of parameters | 27 | 33 |
| $R_{wp}$ | 0.035 | 0.092 |
| GOF | 0.104 | 6.12 |

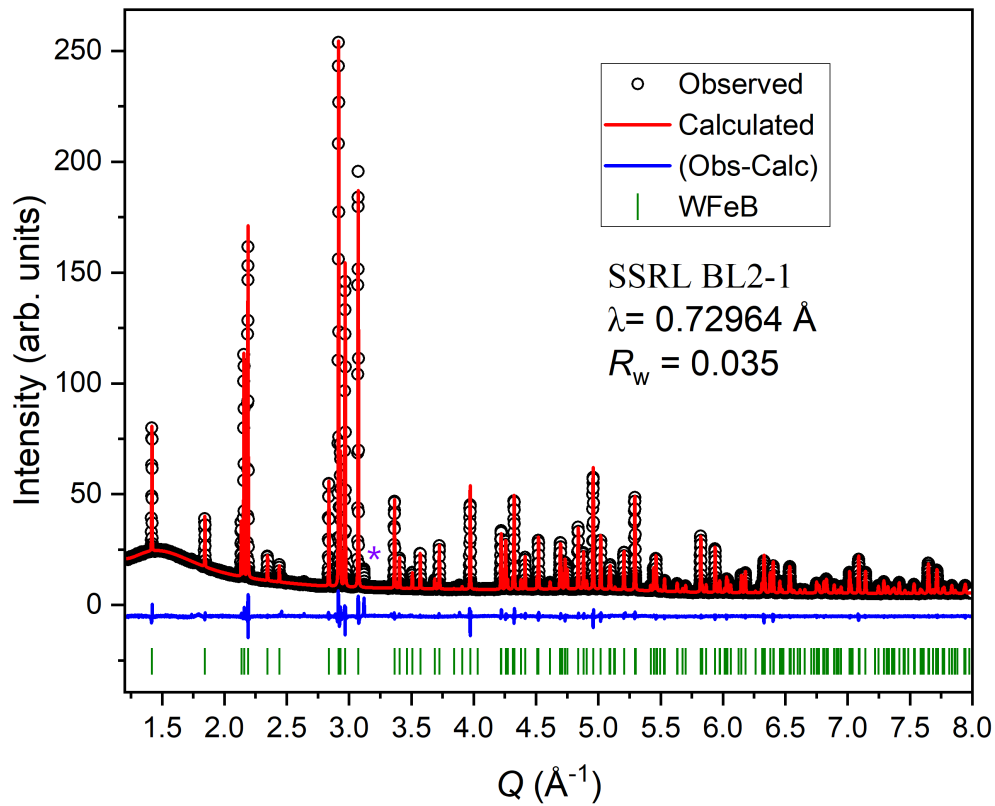

**Figure S2.** Rietveld refinement performed on the synchrotron HR-PXRD ($\lambda$ = 0.72964 Å) data of orthorhombic (*Pnma*) WFeB phase collected at 300 K. The most intense peak of $Fe_2B$ is indicated in a purple asterisk.

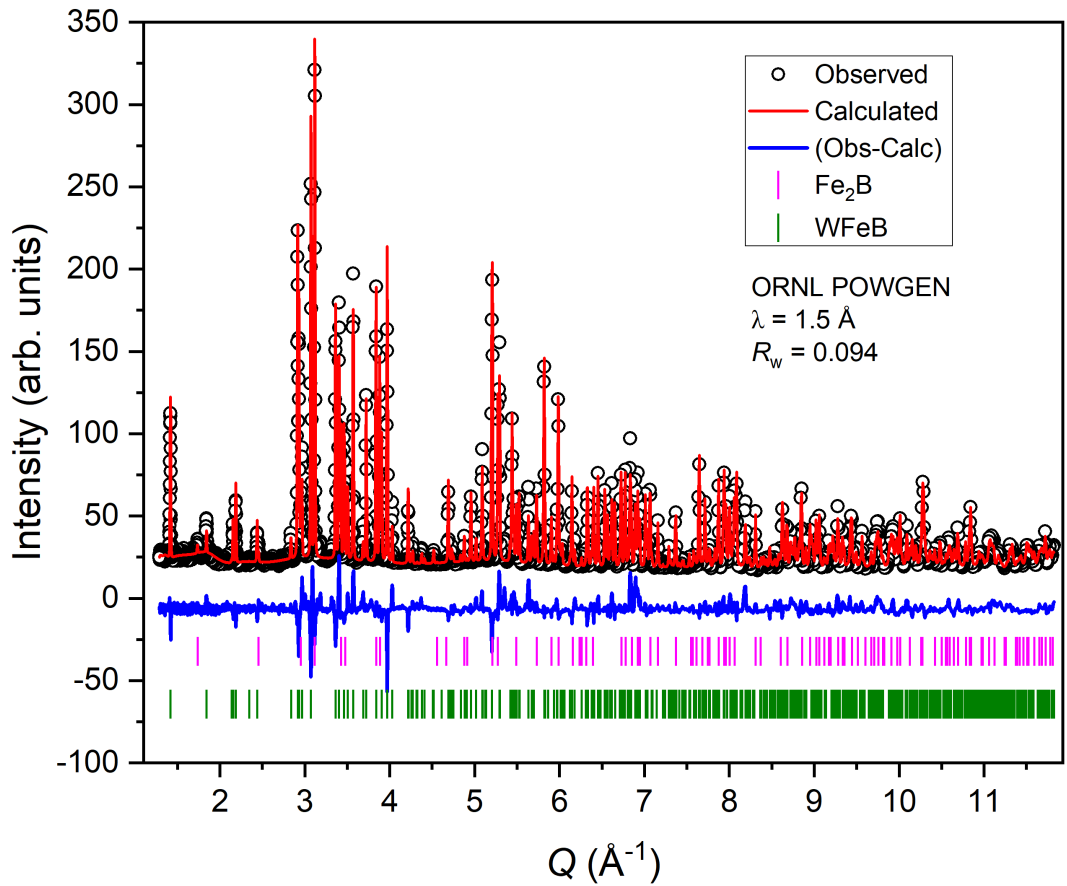

**Figure S3.** Refinement plot of 300 K neutron data ($\lambda$ = 1.5 Å) with no magnetic peaks. The Fe$_2$B impurity phase was included in the refinement, with a derived phase fraction of 20%.

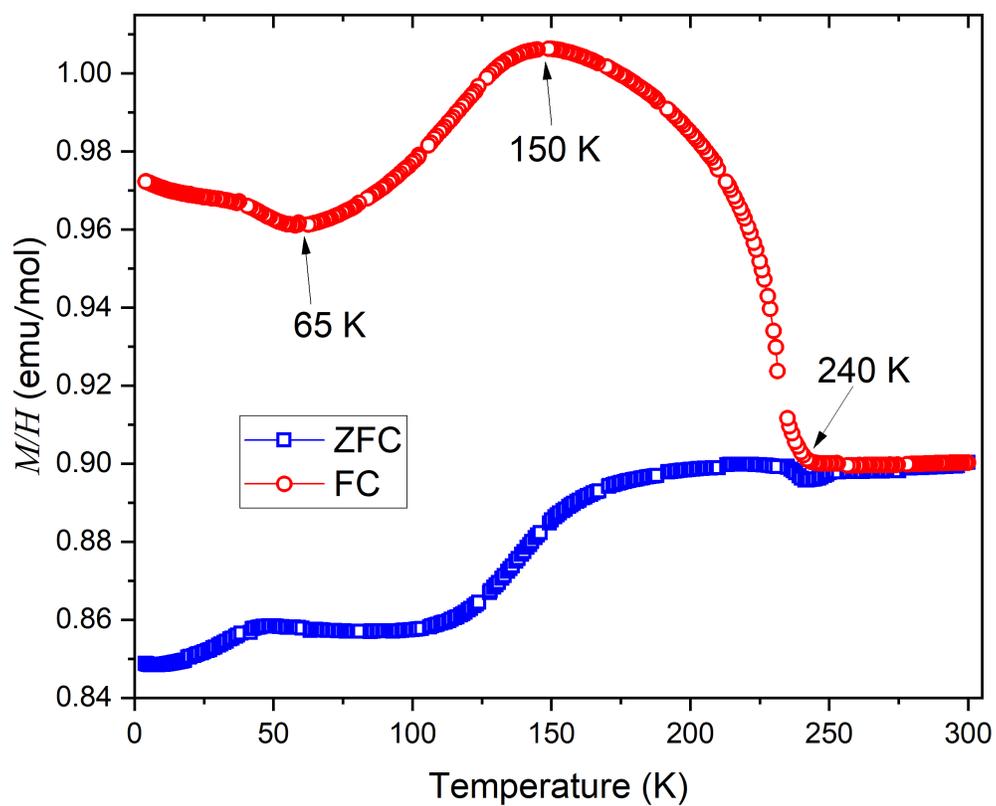

**Figure S4.** *M/H* ZFC-FC as a function of temperature measured under a 1000 Oe applied magnetic field.

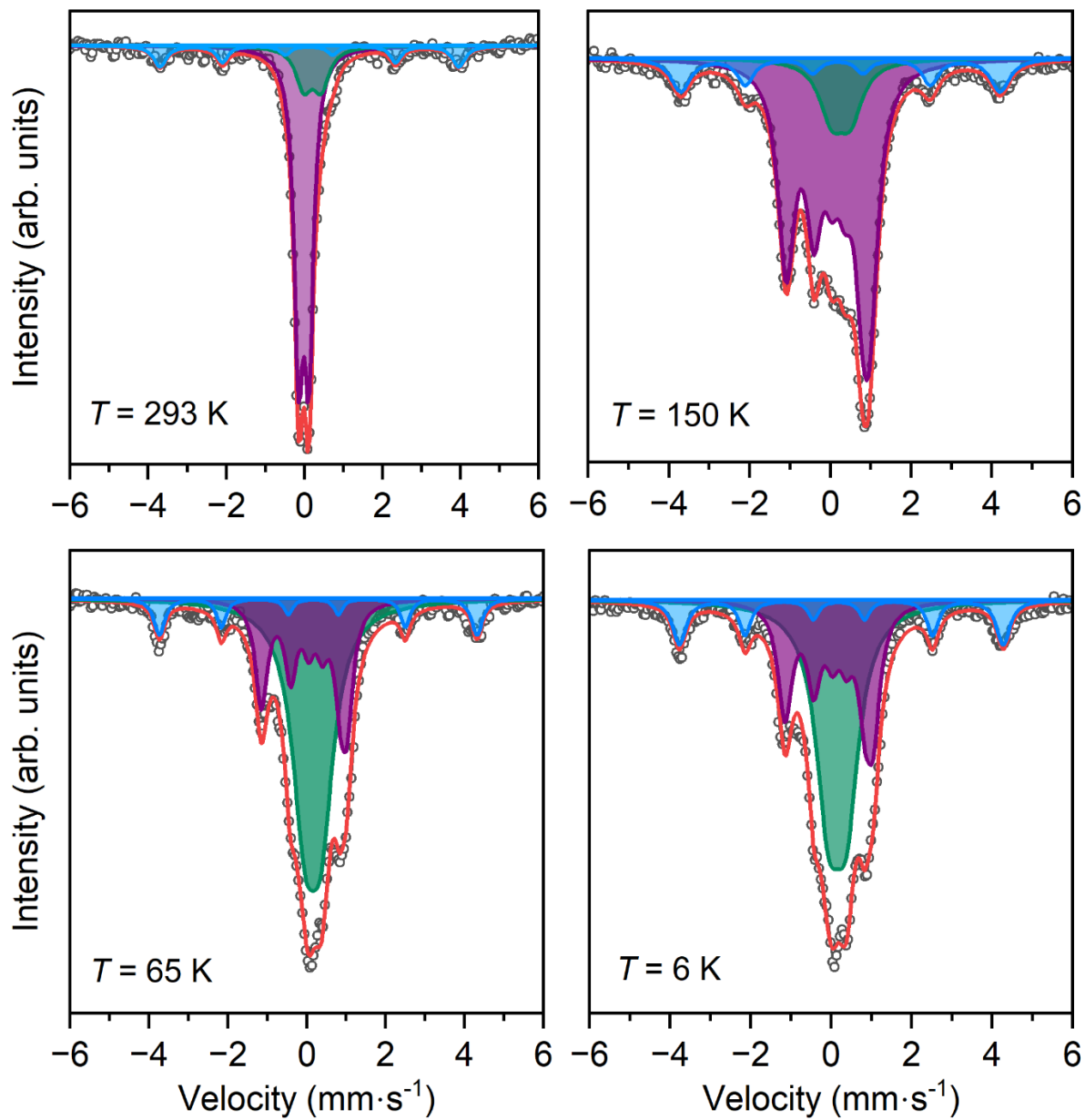

**Figure S5.** $^{57}$Fe Mössbauer spectra of the WFeB sample measured at different temperatures. Experimental pattern: black circles; calculated pattern: red line; $Q_1$ (Fe$_2$B admixture): blue; $Q_2$ (WFeB): purple; $Q_3$ (amorphous admixture): green.

**Table S2.** Summary of refined $^{57}$Fe Mössbauer parameters for the ternary boride sample, WFeB, measured at various temperatures as indicated: centroid shift, δ, quadrupole splitting/quadrupole shift, $\Delta E_Q$/ε, magnetic hyperfine field, $B_{hf}$, full-width at half-maximum, $\Gamma$, and intensity, $I$, of the different components, Q. Estimated errors, except for $I$ being ±2%, are indicated in parenthesis.

| Components | Parameters | 6 K | 65 K | 150 K | 293 K |
|---|---|---|---|---|---|
| Q₁ (Fe₂B) | δ₁ (mm/s) | 0.22(2) | 0.23(2) | 0.22(2) | 0.12(3) |
| | $B_{hf1}$ (T) | 24.9(1) | 25.0(1) | 24.5(2) | 23.7(2) |
| | ε₁ (mm/s) | 0.03(2) | 0.06(2) | 0.03(2) | 0.01(3) |
| | $\Gamma_1$ (T) | 0.31(3) | 0.26(3) | 0.45(4) | 0.31(5) |
| | $I_1$ (%) | 13 | 10 | 14 | 14 |
| Q₂ (WFeB) | δ₂ (mm/s) | 0.07(1) | 0.08(1) | 0.07(1) | -0.02(1) |
| | $B_{hf2}$ (T) | 6.8(1) | 6.7(1) | 6.4(1) | 0 |
| | ε₂/ΔE (mm/s) | 0.13(1) | 0.14(1) | 0.13(1) | 0.27(1)* |
| | $\Gamma_2$ (T) | 0.38(2) | 0.34(2) | 0.46(2) | 0.28(1) |
| | $I_2$ (%) | 36 | 31 | 72 | 69 |
| Q₃ (unknown admixture) | δ₃ (mm/s) | 0.15(1) | 0.15(1) | 0.26(3) | 0.21(6) |
| | ΔE (mm/s) | 0.44(2) | 0.41(2) | 0.46(6) | 0.44(7) |
| | $\Gamma_3$ (mm/s) | 0.76(5) | 0.78(5) | 0.72(8) | 0.49(8) |
| | $I_3$ (%) | 51 | 59 | 14 | 16 |
| Absorption (%) | | 2.5 | 2.8 | 2.3 | 2.9 |

* ΔE value

**Table S3.** Description of the magnetic structure model of WFeB at 8 K, with basic information about its relationship to its parent paramagnetic structure.

| | |
|---|---|
| Parent space group | *Pnma* (# 62) |
| Propagation vector(s) | (0, 0, 0) |
| Transformation from the parent basis to the magnetic lattice | (**a**,**b**,**c**;0,0,0) |
| MSG symbol | *Pnma* (#62.441), |
| Transformation to standard setting of MSG | (a,b,c;0,0,0) |
| Unit cell parameters (Å) | $a$ = 5.8172(1) Å, $b$ =3.15523(6) Å, $c$ = 6.8092 (1) Å, $\alpha$ = 90°, $\beta$ = 90°, $\gamma$ = 90° |
| MSG symmetry operations | 1  x,y,z,+1<br>2  -x,y+1/2,-z,+1<br>3  -x+1/2,-y,z+1/2,+1<br>4  -x+1/2,y+1/2,z+1/2,+1<br>5  -x,-y,-z,+1<br>6  x,-y+1/2,z,+1<br>7  x+1/2,y,-z+1/2,+1<br>8  x+1/2,-y+1/2,-z+1/2,+1 |
| Positions of magnetic atoms (atom, x, y, z) | Fe (4c) (0.1428 0.25000 0.5581) |
| Magnetic moment components ($\mu_B$) of Fe | (x,1/4,z \| 0,$m_y$,0)<br>(-x+1/2,3/4,z+1/2 \| 0,-$m_y$,0)<br>(-x,3/4,-z \| 0,$m_y$,0)<br>(x+1/2,1/4,-z+1/2 \| 0,-$m_y$,0)<br>$m_y$ = 0.8(1) $\mu_B$ |
| Positions of nonmagnetic atoms (atom, x, y, z) | W 0.02876 0.25 0.17118<br>B11 0.26313 0.25 0.86478 |

**Table S4.** Magnetic structure models and magnetic space groups (MSG) that were tested for the (100) single magnetic peak at 150 K. The most different region of NPD at 2.2–2.4 Å$^{-1}$ is highlighted with a pink circle.

***Pnm'a'***
**(#62.447)**
(x,1/4,z | m$_x$,0,m$_z$)
(-x+1/2,3/4,z+1/2 | m$_x$,0,-m$_z$)
(-x,3/4,-z | m$_x$,0,m$_z$)
(x+1/2,1/4,-z+1/2 | m$_x$,0,-m$_z$)

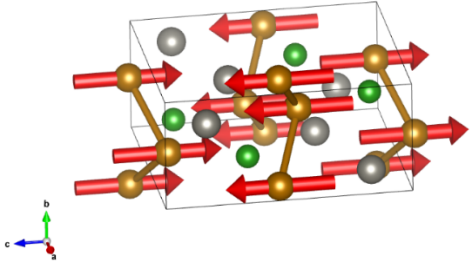
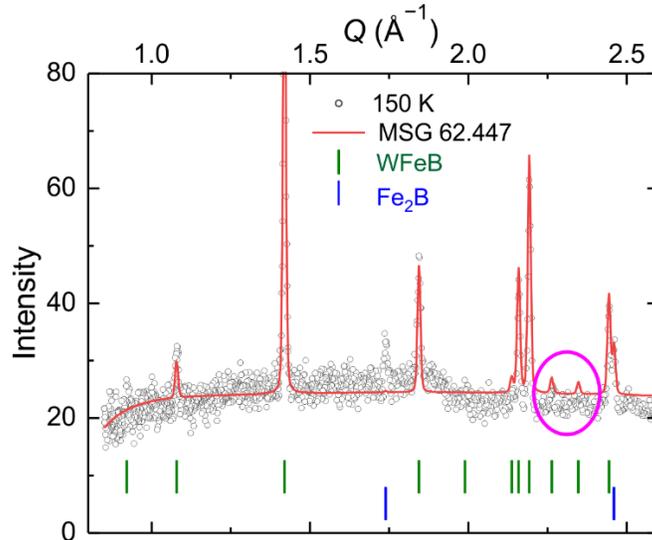

***Pn'm'a'***
**(#62.449)**
(x,1/4,z | m$_x$,0,m$_z$)
(-x+1/2,3/4,z+1/2 | -m$_x$,0,m$_z$)
(-x,3/4,-z | -m$_x$,0,-m$_z$)
(x+1/2,1/4,-z+1/2 | m$_x$,0,-m$_z$)

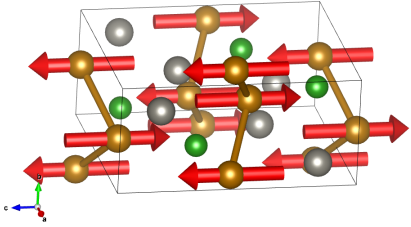
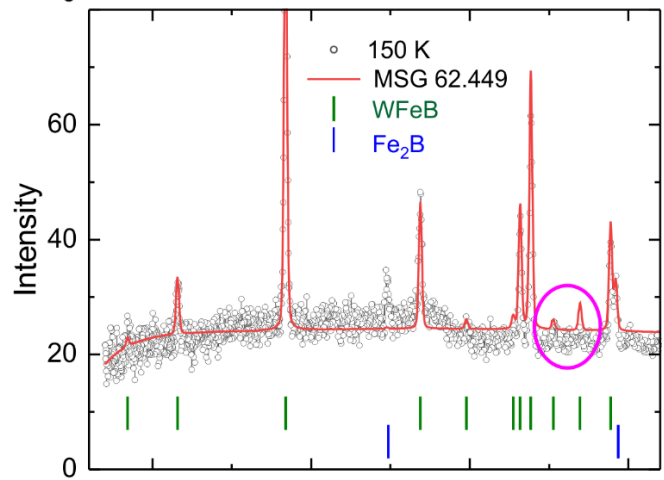

***Pn'ma*** **(#62.443)**
(x,1/4,z | 0,m$_y$,0)
(-x+1/2,3/4,z+1/2 | 0,m$_y$,0)
(-x,3/4,-z | 0,-m$_y$,0)
(x+1/2,1/4,-z+1/2 | 0,-m$_y$,0)

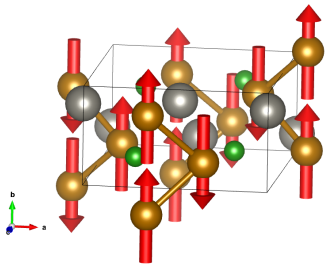
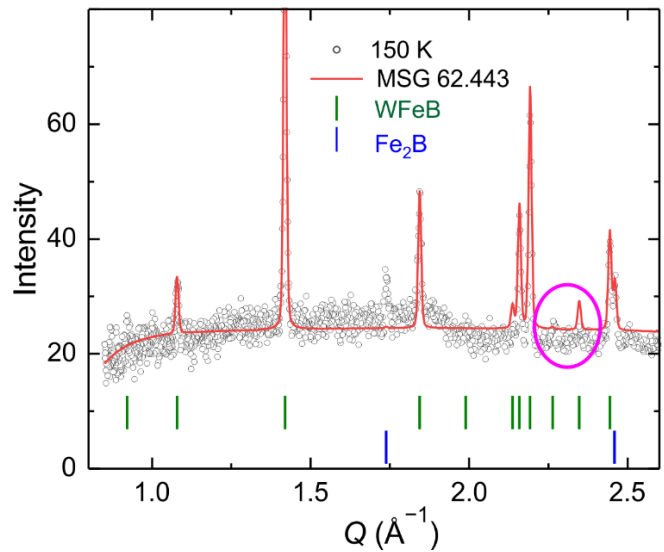

**Table S5.** Description of the magnetic structure (best fit) model of WFeB at 150 K, with basic information about its relationship to its parent paramagnetic structure.

| | |
|---|---|
| Parent space group | *Pnma* (# 62) |
| Propagation vector(s) | (0, 0, 0) |
| Transformation from the parent basis to the magnetic lattice | (**a**,**b**,**c**;0,0,0) |
| MSG symbol | *Pnm'a'* (#62.447), |
| Transformation to standard setting of MSG | (a,b,c;0,0,0) |
| Unit cell parameters (Å) | $a$ = 5.8183 Å, $b$ = 3.1572 Å, $c$ = 6.8100 Å, $\alpha$ = 90°, $\beta$ = 90°, $\gamma$ = 90° |
| MSG symmetry operations | 1  x,y,z,+1 <br> 2  -x,y+1/2,-z,-1 <br> 3  x+1/2,y,-z+1/2,-1 <br> 4  x+1/2,-y+1/2,-z+1/2,+1 <br> 5  -x,-y,-z,+1 <br> 6  x,-y+1/2,z,-1 <br> 7  -x+1/2,-y,z+1/2,-1 <br> 8  -x+1/2,y+1/2,z+1/2,+1 |
| Positions of magnetic atoms (atom, x, y, z) | Fe (4c) (0.1428 0.25000 0.5581) |
| Magnetic moment components ($\mu_B$) of Fe | (x,1/4,z \| $m_x$,0,$m_z$) <br> (-x+1/2,3/4,z+1/2 \| $m_x$,0,-$m_z$) <br> (-x,3/4,-z \| $m_x$,0,$m_z$) <br> (x+1/2,1/4,-z+1/2 \| $m_x$,0,-$m_z$) <br><br> $m_x$ = 0.4(3) $\mu_B$, $m_z$ = 0.7(1) $\mu_B$ |
| Positions of nonmagnetic atoms (atom, x, y, z) | W 0.02876 0.25 0.17118 <br> B11 0.26313 0.25 0.86478 |

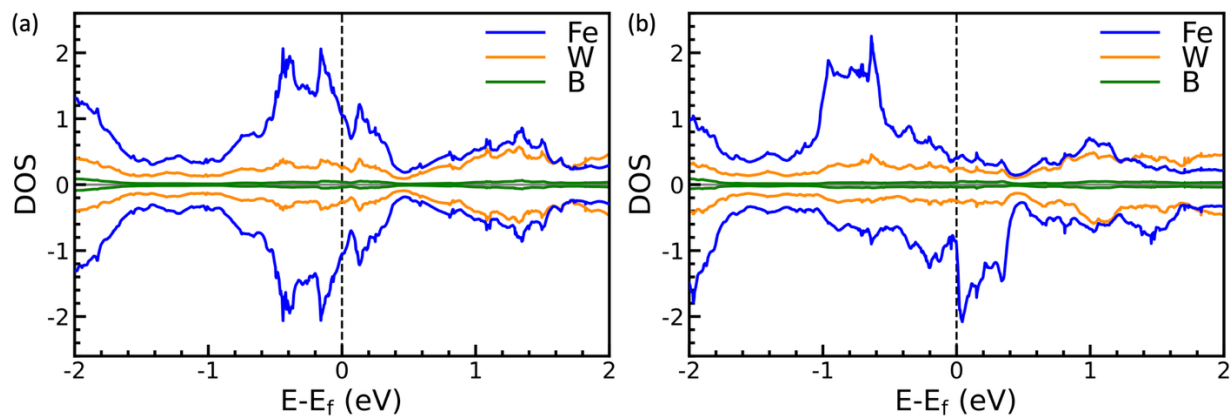

**Figure S6.** (a) Nonmagnetic electronic density of states (DOS) and (b) spin-polarized DOS of the altermagnetic ground state. Blue, orange, and green curves represent the partial DOS (PDOS) of a single Fe, W, and B atom, respectively. Up and down spins represent the majority and minority spins, respectively, for an atom. The unit of the DOS is eV$^{-1}$ atom$^{-1}$ spin$^{-1}$.

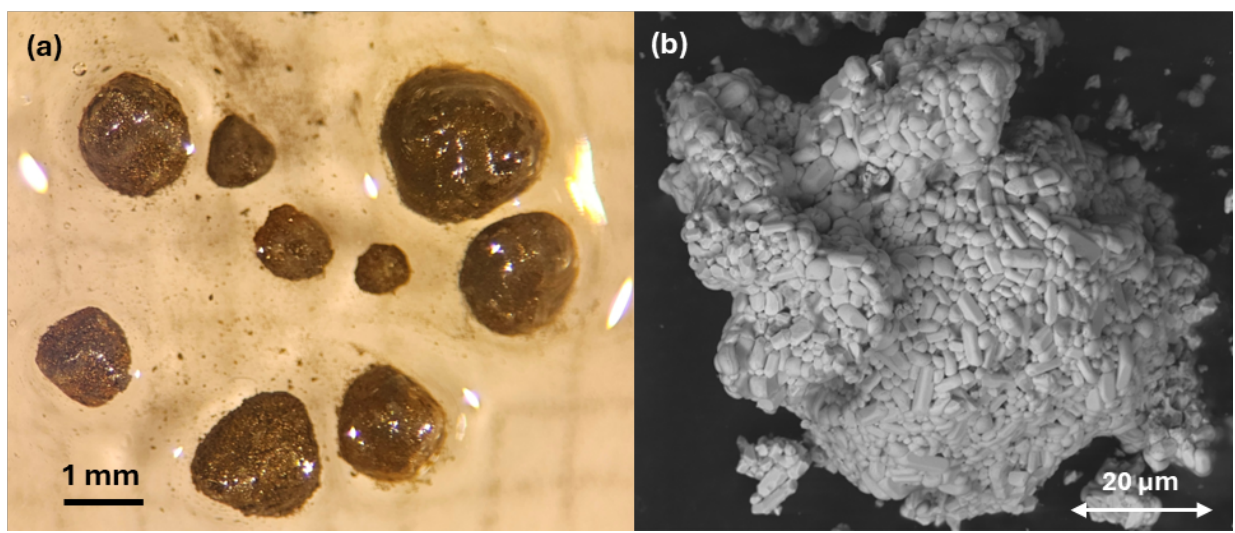

**Figure S7.** (a) Optical microscopy images of WFeB chunks (22.5×). (b) Back-scattered electron images of polycrystalline powders of WFeB.

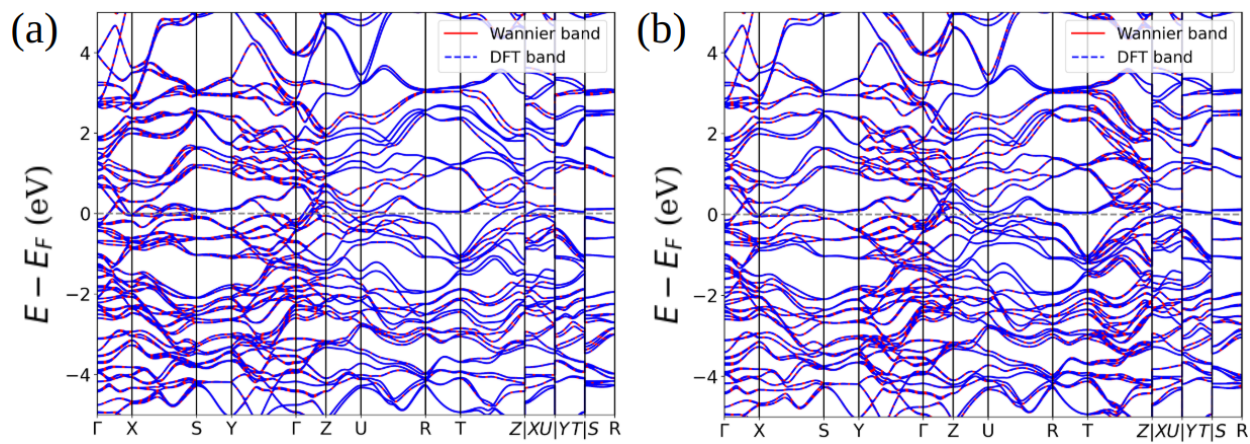

**Figure S8.** Comparison of the electronic band structure of WFeB obtained from first-principles calculations, in the presence of SOC, and the corresponding Wannier-based tight-binding Hamiltonians. (a) Néel vector **L** along the [001] direction. (b) **L** ∥ [100].